\documentclass[12pt,a4paper]{article}

\usepackage[utf8]{inputenc}
\usepackage{amssymb}
\usepackage{multicol}
\usepackage[margin=2.2cm]{geometry}
\usepackage{setspace}
\usepackage{graphicx}
\usepackage{url}
\usepackage{color}
\usepackage{hyperref}
\graphicspath{ {images/} }
\usepackage{tikz}
\usepackage{amsmath}
\usepackage{physics}
\usepackage{ulem}
\usepackage{xcolor}
\usepackage{centernot}
\usepackage{amsthm}

\title{On the Categoricity of Quantum Mechanics}

\author{Iulian D. Toader}

\date{Institute Vienna Circle \\ University of Vienna}

\begin{document}

\doublespacing

\maketitle

\begin{abstract}
    
The paper offers an argument against an intuitive reading of the Stone-von Neumann theorem as a categoricity result, thereby pointing out that, against what is usually taken to be the case, this theorem does not entail any model-theoretical difference between the theories that validate it and those that don't.
\end{abstract}

\bigskip

\section{Introduction}

If quantum mechanics is a categorical theory, we say that it has a unique semantics up to isomorphism. But what does this actually mean? Does it mean a unique interpretation? Since  there are obviously many distinct interpretations, quantum mechanics would have to be considered non-categorical, which would likely strengthen the already formidable battery of arguments for scientific anti-realism. In order to address the categoricity problem for quantum mechanics, we need to first clarify the relevant notions and, in particular, to determine what are the models that should be considered in this context.

There are several historical contexts in which questions concerning the categoricity of quantum mechanics have been raised. Einstein's version of the argument given in the famous EPR paper (Einstein, Podolski, and Rosen 1935), a version that Einstein explained in his letters to Schr\"odinger, has been taken to deploy a notion of incompleteness as non-categoricity (Howard 1990). The suggestion is motivated by the fact that quantum mechanics fails to assign a unique wavefunction to the same physical state of a system, since the assignment depends on the measurement performed on a spacelike separated but previously interacting system. Weyl's criticism of Birkhoff and von Neumann's quantum logic, which they presented in an equally famous paper (Birkhoff and von Neumann 1936), can be understood as emphasizing that this logic is non-categorical, in the sense that it allows non-unique probability valuations to the same experimental propositions (Weyl 1940). Assuming that, as intended by its proponents, the algebraic structure of quantum logic expresses the logical structure of quantum mechanics, this criticism suggests that quantum mechanics is non-categorical as well. On the other hand, the Stone-von Neumann theorem, which establishes the unitary equivalence of the Hilbert space representations of the Weyl algebra defined by canonical commutation relations (Stone 1930, von Neumann 1931), is sometimes read as a categoricity result for quantum mechanics (e.g., St\"oltzner 2002). This variety of contexts easily suggests that one may not always mean the same thing when talking about the categoricity of quantum mechanics. 

A rigorous reconstruction of Einstein's argument, which emphasizes the conditions that justify an understanding of incompleteness as non-categoricity, is given elsewhere (Toader 2020). The key insight is the following: ``if one understands a theoretical state as, in effect, a model for a set of equations plus boundary conditions ..., then Einstein's conception of a completeness requirement should really be understood as a categoricity requirement.'' (Howard 1992, 208) Einstein's argument is sometimes taken to emphasize not the incompleteness, but rather the ``overcompleteness'' of quantum mechanics (Lehner 2014, 319), in the sense that a theory is overcomplete if it allows different theoretical states to describe the same physical state. But whether one calls it incompleteness or overcompleteness, if there are distinct wavefunctions for the same system, and if one is justified to consider them as models, this is enough to show that quantum mechanics is non-categorical. Since space does not allow it here, a detailed analysis of Weyl's argument to the effect that quantum logic is non-categorical is also offered separately (Toader 2020). Weyl showed that there is a one-to-many map between Birkhoff and von Neumann's lattice of experimental propositions and the set of probability values. This fact, he concluded, turned quantum logic into ``the pottage of a nice formal game.'' (Weyl 1940, 299) After Gleason's theorem, arguments for the non-categoricity of quantum logic turned on its connectives not being truth-functional (Hellman 1980) and on its allowing non-isomorphic lattices as models (Pavi\v{c}i\'{c} and Megill 1999).

In the present paper, I will focus on the reading of the Stone-von Neumann theorem as a categoricity result for quantum mechanics. Before I do so, and in order to motivate the main question, I will present the problem of categoricity as discussed in philosophy of mathematics and philosophy of logic, briefly considering some reasons for and some reasons against thinking that categoricity is a desirable property. Turning to physical theories, categoricity seems valuable to scientific realists, especially if they are sensitive to the force of Putnam's model-theoretical argument, but also to some anti-realists, who can argue that even if a theory does not describe physical reality, it can nevertheless be objective, for if it is categorical then the meaning of its concepts is precisely determined. Afterwards, I take up the question about the categoricity of quantum mechanics, first by clearing up some misgivings based on views about the nature of quantum mechanics and its interpretations and reconstructions, and then by offering an argument against the intuitive reading of the Stone-von Neumann theorem as a categoricity result. 

My argument makes good the claim that the relation of unitary equivalence established by this theorem grounds a type of intertranslatability whose semantic counterpart is insufficient for squeezing the models of quantum mechanics into an isomorphism class. I take the view that the models of quantum mechanics are the Hilbert space representations, but I reject the assumption that the isometric isomorphism that underlies their unitary equivalence is the model-theoretical isomorphism required for categoricity. One reason for this is that intertranslatability, understood as definitional equivalence, is a first-order relation, whereas Hilbert space representations cannot be formalized as first-order models. I briefly consider alternative formalizations of representations and different notions of intertranslatability, and I conclude that, against what is usually taken to be the case, the Stone-von Neumann theorem does not entail any model-theoretical difference between the theories that validate it and those that don't.

\section{The Problem of Categoricity}

A theory is said to be categorical with respect to an isomorphism class of models if and only if all its models are in that class. Informally, a model is a set-theoretical structure whose elements can be assigned to the vocabulary of the theory (i.e., the symbols for variables, constants, functions, and relations) by a valuation or interpretation map. The vocabulary, together with the rules for manipulating its symbols, constitute the syntax of the theory, while models constitute its semantics. A categorical theory has a unique semantics up to a suitable isomorphism. 

In philosophy of mathematics and philosophy of logic, the problem of categoricity is discussed especially in connection with arithmetic, analysis, set theory, and classical logic. First-order arithmetic, for example, allows non-standard models (see, e.g., Skolem 1955), and classical logic allows non-normal interpretations, in which disjunction and negation are not truth-functional connectives (Carnap 1943; see also Raatikainen 2008). The problem, as Tarski noted, can be put as follows: ``A non-categorical set of sentences (especially if it is used as an axiom system of a deductive theory) does not give the impression of a closed and organic unity and does not seem to determine precisely the meaning of the concepts contained in it.'' (Tarski 1934, 311) The language of a non-categorical theory is semantically indeterminate, i.e., indeterminate with respect to not only the reference of its terms, but also with respect to the truth value of its sentences: some sentences are true in some models, but false in others. By contrast, a categorical theory determines the meaning of its concepts precisely, and its models agree on all theorems in its language. If ``true'' is understood as ``true in all models'', and ``false'' as ``false in all models'', then categoricity entails bivalence: all sentences in the language are either true or false. The meaning of arithmetical concepts is precisely determined if arithmetic is categorical with respect to the isomorphism class of an omega sequence, that is if it allows only models in which arithmetical vocabulary is used in a natural way, i.e., in agreement with the natural numbers. Similarly, the meaning of classical logical connectives (such as disjunction and negation) is precisely determined if classical logic is categorical with respect to the isomorphism class of a two-element Boolean algebra, that is if it allows only models in which the connectives are used in the normal way, i.e., in agreement with the normal truth tables.

Categoricity appears thus to be considered a desirable property because it blocks semantic indeterminateness by allowing one to pin down the intended unique semantics of a theory, like the natural numbers for arithmetic and the normal truth tables for classical logic. This motivates the attempt to prove theories categorical. But in the case of the most interesting first-order theories, metatheoretical properties like compactness or L\"owenheim-Skolem or G\"odel incompleteness frustrate any such attempt. Therefore, proving categoricity often requires augmenting the language of a theory or adjusting its logic (Corcoran 1980). As is well known, full second-order logic proves the categoricity of arithmetic (Dedekind 1888, Shapiro 1991), although there are strong and well-known arguments against using such a powerful model-theoretical device (Read 1997). As has been often pointed out, second-order quantification is itself semantically indeterminate, so the problem just gets pushed up a level, for in order to pin down the intended model of arithmetic up to isomorphism, one must first be able to pin down the intended model of second order real analysis up to isomorphism (Button and Walsh 2018, 159). It also turns out that attempting to prove the categoricity of arithmetic by means of any logic weaker than second-order logic necessarily presupposes the categoricity of the underlying logic (Read 1997). Considerations like these seem sufficient to make one think that although categoricity appears to be a desirable property, one can never actually pin down the intended unique semantics of a theory without first pinning down the intended unique semantics of another background theory. 

But is categoricity really indispensable? Some argued that the existence of non-standard models of first-order arithmetic is no real concern, because any two non-isomorphic models of it are arithmetically indiscernible, in the sense that the application of induction within the language of the theory yields arithmetical results that hold in any model. Thus, all models of arithmetic agree on all theorems in its language (Resnik 1996). But the fact that provability obliterates the differences between non-isomorphic models does not seem enough to guarantee that one pins down the intended model of arithmetic. For there exist unprovable sentences in the language of the theory on which non-isomorphic models do disagree, which one would have to simply dismiss as surplus structure. It has also been noted that the non-categoricity of other mathematical theories, like the theories of fields, rings, groups, algebras, etc., is a result of design and acumen. Their non-categoricity is not considered a failure, but rather a theoretical success. No intended unique semantics is typically supposed to exist for such theories. Normally, they have a variety of distinct non-isomorphic models, and they may all be thought of as intended. Of course, this does not mean that categoricity, if it can be obtained, would necessarily be a disadvantage. Some categorical algebraic theories have some rather nice features like quantifier elimination. Moreover, semantic indeterminateness does not seem to be considered a problem for such theories. This view, which has been recently extended also to arithmetic and set theory (Hamkins 2012), stems from considerations put forth by some logicians and mathematicians in the early part of the 20th century, who believed that non-categoricity is a theoretical virtue, rather than a liability, and one should formulate axioms that are weak enough to admit a multitude of non-isomorphic models (Zermelo 1930).

The problem of categoricity does not loom large in discussions in philosophy of science, at least by comparison to counterpart discussions in philosophy of mathematics and philosophy of logic. Is categoricity a desirable feature of physical theories? Is it perhaps indispensable? Is the existence of non-isomorphic models of a physical theory a reason to be concerned about the semantic indeterminateness of its language? Categoricity is sometimes rejected by philosophers of science as a kind of ``rigidity or inflexibility'' that should have no place in physics (Bunge 1973). But Einstein and Weyl, as we have noted in the introduction, seem to have thought that non-categoricity was a reason for some serious epistemological concerns (see also Toader 2011 for discussion). Non-categorical theories are sometimes considered more abstract than categorical ones, in the sense that their class of models is too large and diverse to allow a simple mental image of a typical model despite dropping some of its complex properties (Suppes 1993, 74). Suppes seems to have thought that such abstractness should better be avoided for a scientific theory. Putnam's model-theoretical argument, of course, turns non-categoricity (more exactly, the non-categoricity of an ideal first-order theory) into a weapon against metaphysical realism, by making the point that such a theory is vacuously true, i.e., true in any model of any infinite cardinality (Putnam 1980). This argument has fuelled a lot of discussion in the last 40 years that I have no space or intention to review here. But one should note that while no first-order axiomatization has yet been given for quantum theory, such axiomatizations have been offered for special and general relativity (see, e.g., Andr\'eka \textit{et al.} 2012), so one might actually dispense with the assumption of an ideal theory. In any case, it seems that if a physical theory were categorical, this should be good news for scientific realists, who would argue that the theory describes (the structure of) physical reality without having to further justify any choice between its models. But categoricity would presumably be welcomed by anti-realists as well, who could argue that even if a theory does not describe physical reality, it is nevertheless objective, for if it is categorical then the meaning of its concepts is precisely determined by its axioms and rules. Thus, in philosophy of science, categoricity would appear almost generally desirable and perhaps even indispensable (if one is sensitive to the force of Putnam's model-theoretical argument). Indeed, some scholars insist that one should continue to pursue categorical physical theories despite the metatheoretical frustration noted above (Zilber 2016). 

\section{The Categoricity of Quantum Mechanics}

When one turns to quantum mechanics, one finds that even if considered a desirable property, categoricity should be impossible to prove. For instance, as Howard noted, one sometimes argues that if one were to prove the categoricity of a physical theory, this would go against established metatheoretical truth: ``... a corollary of G\"odel's first incompleteness theorem asserts that any theory as powerful as or more powerful than Peano arithmetic -- in first-order formulation -- will be not only deductively incomplete but also non-categorical, which is to say that it will have models that are not isomorphic to one another. And since any moderately sophisticated theory in physics will incorporate a mathematical apparatus as powerful as or more powerful than arithmetic, the same will be true of our physics.'' (Howard 2012) But of course there are consistent theories at least as powerful as first-order arithmetic, which are deductively incomplete but categorical, such as second-order arithmetic. So the argument is valid only if one assumes that theories of physics are first-order. On the other hand, the categoricity of quantum mechanics is sometimes said to have been already proved by von Neumann, as we will see in the next section, where I will analyze in some detail this alleged categoricity theorem.

But before I do so, we should resolve some possible misgivings about the categoricity problem for quantum mechanics. Some may doubt that this a well-posed problem in the first place, because quantum mechanics is not a physical theory. Rather, as has been often contended, it is ``a very effective and accurate \textit{recipe} for making certain sorts of predictions'' (Maudlin 2019, 2). Although one of the most successful achievements of modern science, quantum mechanics is not a theory, but rather ``a general method, a framework in which many theories can be developed'' (Lalo\"e 2019, 13). This very fact allows quantum mechanics to be applied to widely diverse fields way beyond its initial atomic domain. On the face of it, it's hard to see why wide applicability should turn quantum mechanics into something less than a theory. But such claims as just quoted are of course motivated by the view that a physical theory must provide a physical ontology, which means that quantum mechanics can only be considered a theory if taken together with an interpretation. Without an interpretation, quantum mechanics includes just the standard axioms for states and observables, and the Born rule for assigning expectation values to some observables in certain states. These can be formulated as follows: $\psi$ is a state of a system if and only if it is represented by a normed ray (an equivalence class of vectors) of an infinite-dimensional Hilbert space $\mathcal{H}$. \textit{A} is an observable of the system only if it is represented by linear self-adjoint operators acting on $\mathcal{H}$. $\langle A \rangle$ is an expectation value of an observable \textit{A} for a system in a state $\psi$ if and only if it is equal to the inner product $\langle \psi, A\psi \rangle$. A dynamical law (i.e., the Schr\"odinger equation) would also be included to describe the evolution in time of a quantum system. On the other hand, of course, many think that quantum mechanics needs no interpretation (e.g., Fuchs and Perez 2000). Be that as it may, this certainly does not mean that the categoricity problem is not well-posed, since the problem can in fact be raised for any set of axioms, whether you consider it a full-fledged physical theory or not. It may be argued that, in particular, the problem can also be raised for interpretations and reconstructions of quantum mechanics, taken as sets of axioms, but this point will not be pursued here (see Toader 2020 for discussion).

A related misgiving is that even if well-posed, the categoricity problem is really trivial, for the interpretations of quantum mechanics are clearly not isomorphic to one another. Quantum mechanics is a non-categorical theory, admitting models as distinct as, e.g., the pilot wave model, dynamical reduction models, etc., and this is the end of the story. But I think that it's misleading to consider interpretations as models of quantum mechanics. From a logical point of view, most interpretations are extensions of quantum mechanics -- not extensions beyond quantum correlations, of course, but rather extensions of its axioms and rules that add, for example, further dynamical laws. But the extensions of an axiomatic theory are not its models, although they might imply expansions of these models. For similar reasons, the reconstructions of quantum mechanics cannot be taken as its models, either. Unlike interpretations, reconstructions do not extend the theory, but they attempt to derive its axioms and rules from purportedly more fundamental principles, and thereby recover the models of quantum mechanics. Such attempts arguably follow Russell's regressive method: they proceed backwards from quantum mechanics to, e.g., information-theoretical principles, and then forwards from these principles to the axioms and rules of quantum mechanics. Russell's own reconstruction of arithmetic from logical principles was driven by the need to provide a unique semantics for arithmetic, to give numbers a precisely determined meaning, rather than allowing an infinity of models of arithmetical axioms (Russell 1919). The reconstructions of quantum mechanics may similarly be understood as motivated at least in part by the search for a unique semantics.

Now, if the models of quantum mechanics are neither its interpretations, nor its reconstructions, then what are they? A common view in the philosophy of quantum physics is the following: ``Quantum mechanics, we may say, uses the \textit{models} supplied by Hilbert spaces.'' (Hughes 1989, 79) Similarly, and more recently: ``The models of NRQM are Hilbert spaces, along with a suitable subalgebra of the bounded operators on that Hilbert space.'' (Weatherall 2019, 7) I will adopt this view on the models of quantum mechanics for the rest of the paper. The question we face now is the following: If we take Hilbert space representations as the relevant class of models of quantum mechanics, is there an isomorphism that would turn it into an isomorphism class such that quantum mechanics would thereby be proved categorical? The existence of such an isomorphism appears to be confirmed by the Stone-von Neumann theorem, to which I now turn.

\section{The Stone-von Neumann Theorem}

The Stone-von Neumann theorem states that any irreducible faithful Hilbert space representation of the Weyl algebra, which describes a quantum mechanical system (or, more generally, any system with a finite number of degrees of freedom), is uniquely determined up to a unitary transformation (Stone 1930, von Neumann 1931; see Summers 2001 and Rosenberg 2004 for discussion). A Weyl algebra $\mathfrak{A}$ is a C$^{*}$-algebra generated by the Weyl form of the canonical commutation relations. A Hilbert space representation $(\mathcal{H}, \pi)$ is a $^{*}$-homomorphism $\pi : \mathfrak{A} \rightarrow B(\mathcal{H})$, where $B(\mathcal{H})$ is the set of bounded linear self-adjoint operators on a complex Hilbert space $\mathcal{H}$. The representation is irreducible if no (nontrivial) subspace of $\mathcal{H}$ is invariant under the operators in $\pi(\mathfrak{A})$. If $\pi$ is a $^{*}$-isomorphism, then the representation is also faithful. In the representation theory of C$^{*}$-algebras, the Stone-von Neumann theorem entails that any two irreducible faithful representations $(\mathcal{H}_{1}, \pi_{1})$ and $(\mathcal{H}_{2}, \pi_{2})$ of $\mathfrak{A}$ are unitarily equivalent if and only if there is an element $U\in\mathfrak{A}$ which acts as an operator $U:\mathcal{H}_{1}\rightarrow\mathcal{H}_{2}$ such that $UU^{*}=U^{*}U=1$ and $\pi_{1}(A)$ = $U\pi_{2}(A)U^{*}$ for all elements $A\in\mathfrak{A}$. If $\mathcal{H}_{1}$ and $\mathcal{H}_{2}$ are separable, infinite-dimensional spaces, so their orthonormal bases are both countably infinite, $U$ intertwines all bounded linear self-adjoint operators on the two spaces. Thus, there is an isometric isomorphism between $\mathcal{H}_{1}$ and $\mathcal{H}_{2}$ that underlies the unitary equivalence of $(\mathcal{H}_{1}, \pi_{1})$ and $(\mathcal{H}_{2}, \pi_{2})$.

Unitary equivalence is typically taken to entail physical equivalence, in the sense that the quantum states described as density matrices in unitarily equivalent representations assign the same expectation values to corresponding physical observables (Weyl 1930, 407; Ruetsche 2011, 24sq). This is, for example, the sense in which one has come to speak of the physical equivalence of the Schr\"odinger and the Heisenberg representations of a quantum mechanical system: time evolution on states, in the Schr\"odinger representation, is the same as time evolution on observables, in the Heisenberg representation (see Muller 1997 and Perovic 2008 for discussion of the sense and scope of the equivalence before von Neumann's 1931 proof of unitary equivalence). For this reason, as mentioned above, the Stone-von Neumann theorem has been intuitively interpreted as a categoricity result: ``It happens only in rare cases that all models of an axiom system are isomorphic, so that the axioms and every single model have the same physical content and no subsidiary condition is needed. Even the axioms of ordinary arithmetic admit a non-standard model entirely different from the natural numbers, the so-called Skolem functions. Von Neumann, however, succeeded in proving that quantum \textit{mechanics} has the particularly nice feature -- which logicians call categoricity -- that all representations of the canonical commutation relations are unique up to isomorphism.'' (St\"oltzner 2002, 45) Similarly, and more recently: “[The Stone-von Neumann] theorem can be naturally read as a categoricity result.'' (Toader 2018, 21) Unfortunately, as I will presently argue, this intuitive reading does not stand up to closer scrutiny.

One should first note that, while he considered the categoricity of set theory a desirable property, albeit one rather impossible to obtain (von Neumann 1925, 412), von Neumann never considered the result just described as one that established the categoricity of quantum mechanics. When he later developed his axiomatic theory of games, von Neumann expressly noted that this is intended as non-categorical, since it is meant to describe many different games, rather than a unique class of isomorphic ones, and he compared game theory in this respect with group theory and rational mechanics, which he took to similarly describe many different groups and many different mechanical systems, respectively (von Neumann and Morgenstern 1944, section 10.2). In any case, that the Stone-von Neumann theorem \textit{cannot} be taken as a categoricity result will be argued as follows: unitary equivalence entails the intertranslatability of Hilbert space representations, but the semantic counterpart of this is insufficient for squeezing the representations into an isomorphism class of models, and therefore insufficient for categoricity. This is because intertranslatability, understood as definitional equivalence, is a first-order relation, whereas Hilbert space representations cannot be formalized as first-order models.

I will consider in due time some adjustments to this argument, but first let's develop it in a bit more detail, starting with the claim that the Heisenberg representation and the Schroedinger representation of the Weyl algebra of a quantum system are intertranslatable. What does this claim mean? As Ruetsche put it, this means that ``Any data which elements of the [one representation] accommodate, counterpart elements of the [other representation] accommodate as well – and \textit{mutatis mutandis} for falsifying data.'' (Ruetsche 2011, 45) Intertranslatability is here understood in the more general sense articulated by Glymour: ``[The intertwiner of representations] provides the translation manual [Glymour] is after.'' (\textit{loc. cit.}) Glymour's translation manual is offered as a means for translating between theories, rather than between Hilbert space representations, and it ``guarantees that all and only theorems of [one theory] are translated as theorems of [another theory], and conversely.'' (Glymour 1970, 279) But the data accommodated by unitarily equivalent representations are the expectation values assigned to corresponding physical observables in states described as density matrices in those representations. Therefore, one can understand intertranslatability of representations in Glymour's general sense only if one takes, by analogy, the statements about expectation values as theorems (Clifton and Halvorson 2001, 430). Representations \textit{qua} theories are intertranslatable in the sense that they agree on all theorems. More exactly, and keeping close to Glymour's account, they are intertranslatable if and only if they have a ``common definitional extension'' (Glymour 1970, 279), that is they have definitional extensions that derive the same theorems. But since, following Hughes, we consider Hilbert space representations as models of quantum mechanics, the semantic counterpart of intertranslatability holds if and only if the models have definitional expansions that satisfy the same theorems. If ``true'' is understood as ``true in all models (or their definitional expansions)'', and ``false'' as ``false in all models (or their definitional expansions)'', then unitary equivalence entails bivalence: all statements about expectation values are either true or false. But of course bivalence would be expected, if quantum mechanics were categorical, i.e., if there existed a model-theoretical isomorphism between Hilbert space representations. Unfortunately, without further adjustments, the converse does not hold, for non-isomorphic structures can satisfy the same theorems.

Furthermore, and more importantly, intertranslatability understood as definitional equivalence is a first-order relation, which means that its semantic counterpart holds between Hilbert space representations \textit{qua} models only in the sense that they satisfy all and only first-order theorems. But $(\mathcal{H}_{1}, \pi_{1})$ and $(\mathcal{H}_{2}, \pi_{2})$ are not first-order models, on account of their metric completeness, a property that cannot be expressed in a first-order language. However, fixing this may require only a small adjustment of the background logic, which is worth considering here even if only briefly. Consider continuous first-order logic, which is an extension of first-order logic whose vocabulary consists of constant symbols, \textit{n}-ary function symbols and \textit{n}-ary relation symbols, and which replaces the set of possible truth values from \{T, F\} to the bounded interval [0, 1], and Boolean functions as connectives by continuous functions from [0, 1]$^{n}$ to [0, 1] (Ben  Yaacov \textit{et al.} 2008). Negation, disjunction, conjunction, and implication are defined as follows: $\neg a := 1-a $; $a \vee b := min(a,b)$; $a \wedge b := max(a,b)$; $a \rightarrow b := min(1-a,b)$. First-order quantifiers $\forall x$ and $\exists x$ are replaced by the operations $sup_{x}$ and $inf_{x}$, respectively, which are defined on the complete linear ordering of elements in the unit interval. Continuous first-order logic allows, for example, a formalization of C$^{*}$-algebras (Farah \textit{et al}. 2018). Following this line, suppose one formalizes Hilbert space representations as continuous first-order models of quantum mechanics and suppose that unitary equivalence can also be expressed in this extended language. Might (an appropriately formulated version of) the Stone-von Neumann theorem be rigorously interpreted as a categoricity result for quantum mechanics in this model-theoretical framework? Unfortunately, this cannot be the case: continuous first-order logic is characterized by compactness and L\"owenheim-Skolem properties (Ben  Yaacov \textit{et al.} 2008, see also Baldwin 2018, 97). Thus, $(\mathcal{H}_{1}, \pi_{1})$ and $(\mathcal{H}_{2}, \pi_{2})$ cannot be properly formalized as first-order models, and when formalized as continuous first-order models, they cannot be proved isomorphic. Therefore, quantum mechanics cannot be categorical in continuous first-order logic.

A natural move at this junction is to go even higher in the logical hierarchy, and formalize Hilbert space representations as second-order models of quantum mechanics. Full second-order logic is not characterized by either compactness or L\"owenheim-Skolem properties. However, just as in the case of arithmetic, this would not help much if it only pushed the problem up to the level of second-order quantification.  Furthermore, second-order quantification is often believed to take ontological commitment beyond what would be acceptable -- in the case of physical theories like quantum mechanics, beyond physical reality. One might  resist this point (typically attributed to Quine, although first made by Weyl) and adopt an ontologically deflationary account of second-order quantification, if a suitable one were available for physics. But this would not fix the semantic indeterminateness of second order quantification. Alternatively, it  might seem better to adjust the notion of intertranslatability. Putnam, for example, recently suggested the following: ``The different `representations' are perfectly intertranslatable; it is just that the translations do not preserve `ontology'. What do they preserve? Well, they do not merely preserve macro-observables. They also preserve \textit{explanations}. An explanation of a phenomenon goes over into another perfectly good explanation of the same phenomenon under these translations.'' (Putnam 2015, 23) On this suggestion, unitarily equivalent representations are intertranslatable in the sense that they agree on all explanations. Is this explanatory equivalence sufficient for categoricity? It's hard to say, but as Putnam acknowledged, one would need to develop an account of what an explanation is supposed to be in this context: ``But who's to say what is a phenomenon? And who's to say what is a perfectly good explanation? My answer has always been: \textit{physicists} are not linguists and not philosophers.'' (\textit{loc. cit.})

Another notion of intertranslatability could be articulated in terms of  Morita equivalence, such that a semantic counterpart of it would hold between models of quantum mechanics. As Rosenberg noted, ``The `modern' approach to the Stone-von Neumann Theorem, which is somewhat more algebraic, is due to Rieffel [1972, 1974]. The key observation of Rieffel is that the theorem is really about an equivalence of categories of representations, or in the language of ring theory, a Morita equivalence.'' (Rosenberg 2004, 342) The Stone-von Neumann theorem, as formulated by Rieffel, states the following: ``Every irreducible Heisenberg G-module is unitarily equivalent to the Schr\"odinger G-module.'' (Rieffel 1972) Unitary equivalence is constructed as the Morita equivalence of G-modules, where the G-modules are Hilbert space representations together with the collection of all intertwining operators between them. Then, the semantic counterpart of Morita equivalence holds between G-modules considered as models if and only if they have Morita expansions that satisfy the same theorems. If ``true'' is understood as ``true in all models (or their Morita expansions)'', and ``false'' as ``false in all models (or their Morita expansions)'', then unitary equivalence entails bivalence: all statements about expectation values are either true or false. But again, bivalence would be expected, if quantum mechanics were categorical. As already mentioned, however, the converse does not hold. Moreover, Morita equivalence is a generalized definitional equivalence relation (Barrett and Halvorson 2016), and thus it is first-order, which makes it inadequate for present purposes, since just like the Hilbert space representations, Rieffel's G-modules cannot be formalized as first-order models.

All the above considerations strongly suggest that the intuitive reading of the Stone-von Neumann theorem as a categoricity result is not justified, which makes one doubt that quantum mechanics has a unique semantics up to isomorphism. But as in the case of first-order arithmetic, one could respond by arguing that the categoricity of quantum mechanics is not really indispensable. One could thus emphasize that even if the unitary equivalence of Hilbert space representations fails to squeeze them into an isomorphism class of models, it does show that they are physically indiscernible, in the sense articulated by definitional equivalence -- they agree on all statements about expectation values. But does this agreement obliterate all physical differences between the models of quantum mechanics? This depends on whether unbounded operators, which are not discernible by unitary equivalence, are considered physically significant or dismissed as surplus mathematical structure. If unbounded operators are taken to be physically significant, then not only are they a source of incompleteness in quantum mechanics (Heathcote 1990), but a new notion of intertranslatability would be needed to justify the physical equivalence of Hilbert space representations (Baker 2011, 147).

Alternatively, one may be inclined to adopt the attitude that logicians and mathematicians from Zermelo to Hamkins have adopted in the case of set theory and arithmetic, and appreciate non-categoricity as a theoretical virtue of quantum mechanics. Indeed, following von Neumann's view of axiomatic game theory and rational mechanics, one sometimes speaks, very aptly, of ``intended non-categoricity'' to characterize the intention to construct physical theories that allow for multiple non-isomorphic models (R\'edei 2014, 80). But this intention is typically exemplified by theories like quantum field theory, which invalidate the Stone-von Neumann theorem. This suggests that what is really meant to be described here is an intention to construct theories that allow for unitarily inequivalent representations, so that, as in quantum field theory, one can have representations of the C$^{*}$-algebra on a free field, but also representations on an interacting field. The failure of the Stone-von Neumann theorem for quantum systems with an infinite number of degrees of freedom has been correctly considered sufficient to establish the non-categoricity of a theory describing such systems, like quantum statistical mechanics (in the thermodynamic limit) and quantum field theory: ``Our best current theory is QFT. It is a relativistic theory (in the sense of special, not general relativity), and it is a theory of systems with an infinite number of degrees of freedom. As such, in its most natural algebraic form, it can be shown to possess representations that are, of necessity, unitarily inequivalent. This is the algebraist's way of saying that the theory is not categorical, that it does not constrain the class of its models up to the point of isomorphism.'' (Howard 2011, 231) If my argument about quantum mechanics is sound, and the Stone-von Neumann theorem should not be read as a categoricity result, then the validity of this theorem does not make any model-theoretical difference between quantum mechanics and quantum field theory.

\section{Conclusion}

I have argued against an intuitive reading of the Stone-von Neumann theorem as a categoricity result for quantum mechanics, by pointing out that if one takes Hilbert space representations as its models, the unitary equivalence established by the theorem is not enough for a model-theoretical isomorphism between these models. If this is correct, then it entails that Tarski's characterization of non-categorical sets of axioms may apply to quantum mechanics as well: its language may be semantically indeterminate. It also suggests, more generally, that uniqueness theorems in the representation theory of algebra are not easily assimilated to categoricity results in model theory. Within philosophy of science, this might concern the scientific realist who is sensitive enough to the force of Putnam's model-theoretical argument but cannot find comfort in the physical indiscernibility of Hilbert space representations guaranteed by the Stone-von Neumann theorem. It might also concern the anti-realist who is inclined to take semantic determinateness as a requirement for scientific objectivity. Less concerned will be the valiant who seek reformulations of the Stone-von Neumann theorem in powerful model-theoretical frameworks like full second-order logic, and the nonchalant who see non-categoricity as a theoretical benefit, rather than a semantic fiasco. Such attitudes ought to be carefully developed and analyzed.

\section{Acknowledgements}

I am indebted to Don Howard, Cristi Stoica, and Boris Zilber, as well as audiences in Salzburg and Geneva.

\section{Bibliography}

Andr\'eka, H., J. X. Madar\'asz, I. N\'emeti, and G. Sz\'ekely (2012) ``A logic road from special relativity to general relativity'' in \textit{Synthese}, 186, 633–-649.

Baker, D. (2011) ``Broken Symmetry and Spacetime'', in \textit{Philosophy of Science}, 78, 128--148.

Baldwin, J. (2018) \textit{Model Theory and the Philosophy of Mathematical Practice: Formalization without Foundationalism}, Cambridge University Press.
 
Barrett and H. Halvorson (2016) ``Morita equivalence'' in \textit{The Review of Symbolic Logic}, 9, 556--582. 

Ben Yaacov, I., A. Berenstein, C.W. Henson, \& A. Usvyatsov (2008) ``Model theory for metric structures'', in \textit{Model theory with applications to algebra and analysis}, Cambridge University Press, 315--427.

Birkhoff, G. and J. von Neumann (1936) ``The Logic of Quantum Mechanics'', in \textit{Annals of Mathematics} 37, 823--843.

Bunge, M. (1973) \textit{Philosophy of Physics}, D. Reidel.

Button, T. and S. Walsh (2018) \textit{Philosophy and Model Theory}, Oxford University Press.

Carnap, R. (1943) \textit{Formalization of Logic}, Harvard University Press.

Clifton, R. and H. Halvorson (2001) ``Are Rindler Quanta Real? Inequivalent Particle Concepts in Quantum Field Theory'', in \textit{The British Journal for the Philosophy of Science}, 52, 417--470.

Corcoran, J. (1980) ``Categoricity'' in \textit{History and Philosophy of Logic}, 1, 187--207.

Dedekind, R. (1888) \textit{Was sind und was sollen die Zahlen?}, Braunschweig, Vieweg.

Einstein, A., B. Podolsky, and N. Rosen (1935) ``Can Quantum-Mechanical Description of Physical Reality Be Considered Complete?'', in \textit{Physical Review}, 47, 777--780.

Farah, I., B. Hart, M. Lupini, L. Robert, A. Tikuisis, A. Vignati, \& W. Winter (2018) ``Model theory of C$^{*}$-algebras'', available at arxiv.org/abs/1602.08072

Fuchs, C. A. and A. Perez (2000) ``Quantum Theory Needs No `Interpretation''', in \textit{Physics Today} 53, 3, 70. doi:10.1063/1.883004

Glymour, C. (1970) ``Theoretical realism and theoretical equivalence'', in R. Buck and R. Cohen (eds.) \textit{Boston Studies in Philosophy of Science}, 7, 1971, Dordrecht, Reidel, 275–-288.

Hamkins, J. D. (2012) ``The Set-Theoretic Multiverse”, in \textit{The Review of Symbolic Logic}, 5/3.

Heathcote, A. (1990) ``Unbounded Operators and the Incompleteness of Quantum Mechanics'', in \textit{Philosophy of Science}, 57, 523--534.

Hellman, G. (1980) ``Quantum logic and meaning'', in \textit{Proceedings of the Philosophy of Science Association}, 493--511.

Howard, D. (1990) ```Nicht Sein Kann Was Nicht Sein Darf', Or the Prehistory of the EPR, 1909-1935: Einstein's Early Worries About the Quantum Mechanics of Composite Systems'' in \textit{Sixty-Two Years of Uncertainty: Historical, Philosophical, and Physical Inquiries into the Foundations of Quantum Mechanics}, ed. by A. I. Miller, Plenum Press, 61--111.

Howard, D. (1992) ``Einstein and \textit{Eindeutigkeit}: A Neglected Theme in the Philosophical Background to General Relativity'', in \textit{Studies in the History of General Relativity}, ed. by J. Eisenstaedt and A. J. Kox, Birkhäuser, 154--243. 

Howard, D. (2011) ``The physics and metaphysics of identity and individuality'', in \textit{Metascience}, 20, 225--231.

Howard, D. (2012) ``The Trouble with Metaphysics'', unpublished manuscript. 

Hughes, R. I. G. (1989) \textit{The Structure and Interpretation of Quantum Mechanics}, Harvard University Press. 

Lalo\"e, F. (2019) \textit{Do We Really Understand Quantum Mechanics?}, Cambridge University Press, 2nd ed.

Lehner, Ch. (2014) ``Einstein's Realism and His Critique of Quantum Mechanics'', in \textit{The Cambridge Companion to Einstein}, ed. by M. Janssen and Ch. Lehner, Cambridge University Press, 306--353.

Maudlin, T. (2019) \textit{Philosophy of Physics: Quantum Theory}, Princeton University Press. 

Muller, F. A. (1997) ``The equivalence myth of quantum mechanics'', \textit{Studies in History and Philosophy of Modern Physics} 28, 35--61 and 219--247.

Pavi\v{c}i\'{c}, M. and N. D. Megill (1999) ``Non-orthomodular models for both standard quantum logic and standard classical logic: repercussions for quantum computers'', in \textit{Helvetica Physica Acta}, 189--210.

Perovic, S. (2008) ``Why were Matrix Mechanics and Wave Mechanics considered equivalent?'', \textit{Studies in History and Philosophy of Modern Physics}, 39, 444--461.

Putnam, H. (1980) ``Models and Reality'', in \emph{The Journal of Symbolic Logic}, 45, 464--482.

Putnam, H. (2015) ``From quantum mechanics to ethics and back again'', in \textit{Reading Putnam}, ed. by M. Baghramian, Routledge, 19--36.

Raatikainen, P. (2008) ``On rules of inference and the meanings of logical constants'', in \emph{Analysis}, 68, 282--287.

Read, S. (1997) ``Completeness and Categoricity: Frege, G\"odel and Model Theory'' in \textit{History and Philosophy of Logic}, 18, 79--93.

R\'edei, M. (2014) ``Hilbert's 6th Problem and Axiomatic Quantum Field Theory'', in \textit{Perspectives on Science}, 22, 80--97. 

Resnik, M. (1996) ``Structural Relativity'', in \textit{Philosophia Mathematica}, 3, 83--99. 

Rieffel, (1972) ``On the Uniqueness of the Heisenberg Commutation Relations'', in \textit{Duke Mathematical Journal}, 39, 745--752.

Rosenberg, J. (2004) ``A selective history of the Stone–von Neumann Theorem'', in \textit{Operator Algebras, Quantization, and Noncommutative Geometry}, Providence, RI, American Mathematical Society, 331–-354.

Ruetsche, L. (2011) \textit{Interpreting Quantum Theories. The Art of the Possible}, Oxford University Press.

Russell, B. (1919) \textit{Introduction to Mathematical Philosophy}, George Allen \& Unwin, Ltd., London.

Shapiro, S. (1991) \emph{Foundations Without Foundationalism: A Case for Second-Order Logic}, Oxford University Press. 

Skolem, Th. (1955) ``Peano's Axioms and Models of Arithmetic'', in \textit{Studies in Logic and the Foundations of Mathematics}, 16, 1--14.

Stone, M. H. (1930) ``Linear Transformations in Hilbert Space'', in \textit{Proceedings of the National Academy of Sciences of the USA}, 16, 172--175.

St\"oltzner, M. (2002) ``Bell, Bohm, and Von Neumann: Some Philosophical Inequalities Concerning No-Go Theorems and the Axiomatic Method'', in \textit{Non-locality and Modality}, ed. by T. Placek and J. Butterfield, Springer, 37--58.

Summers, S. J. (2001) ``On the Stone-von Neumann Uniqueness Theorem and Its Ramifications'', in \textit{John von Neumann and the Foundations of Quantum Physics}, ed. by M. R\'edei and M. St\"oltzner, Springer, 135--152.

Suppes, P. (1993)\textit{Models and Methods in the Philosophy of Science: Selected Essays}, Dordrecht, Springer.

Tarski, A. (1934) ``Some methodological investigations on the definability of concepts'', in his \textit{Logic, Semantics, Metamathematics}, Hackett, 298--319.

Toader, I. D. (2011) \textit{Objectivity Sans Intelligibility: Hermann Weyl’s Symbolic
Constructivism}, PhD dissertation, University of Notre Dame. Published by \textit{Pro-Quest}, University of Michigan, 2012.

Toader, I. D. (2018) ```Above the Slough of Despond': Weylean Invariantism and Quantum Physics'', in \textit{Studies in History and Philosophy of Modern Physics}, 61, 18--24.

Toader, I. D. (2020) \textit{The Categoricity of Quantum Theory}, unpublished manuscript.

von Neumann, J. (1925) ``An axiomatization of set theory'' in \textit{From Frege to G\"{o}del: A Source Book in Mathematical Logic}, 1879-1931, ed. by J. van Heijenoort, Harvard University Press, 1967, 393--413.

von Neumann, J. (1931) ``Die Eindeutigkeit der Schr\"{o}dingerschen Operatoren'', in \emph{Mathematische Annalen}, 104, 570--578.

von Neumann, J. and O. Morgenstern (1944) \textit{Theory of Games and Economic Behavior}, Princeton University Press.

Weatherall, J. O. (2019) ``Part 1: Theoretical equivalence in physics'', in \textit{Philosophy Compass}, doi:10.1111/phc3.12592

Weyl, H. (1930) \textit{Gruppentheorie und Quantenmechanik}, 2nd ed.,
Leipzig, Hirzel. Eng. tr. as \textit{The Theory of Groups and Quantum Mechanics}, Dover, 1931.

Weyl, H. (1940) ``The Ghost of Modality'', in \textit{Philosophical Essays in Memory of Edmund Husserl}, ed. by M. Farber, Cambridge University Press, 278--303.

Zermelo, E. (1930) ``\"{U}ber Grenzzahlen und Mengenbereiche. Neue Untersuchungen \"{u}ber die Grundlagen der Mengenlehre'', in \emph{Fundamenta Mathematicae}, 16, 29-47. Reprinted with an English translation in \emph{Collected Works. Vol. I}, ed. by H.-D. Ebbinghaus and A. Kanamori, Springer, 400-430.

Zilber, B. (2016) ``The semantics of the canonical commutation relations'', available at arxiv.org/abs/1604.07745

\end{document}